# Interaction between vegetation and Snowball phases in the late Proterozoic Earth

Erica Bisesi[1,2,*], Giuseppe Murante[1,2,3,4], Antonello Provenzale[2], Jost von Hardenberg[5,6], Michele Maris[1,3,4] and Laura Silva[1,3]

[1]INAF − Astronomical Observatory of Trieste, Via G. Tiepolo 11, 34143 Trieste, Italy; [2]CNR – Institute of Geosciences and Earth Resources, Via G. Moruzzi 1, 56124 Pisa, Italy; [3]Institute for Fundamental Physics of the Universe, Via Beirut 2, 34151 Trieste, Italy; [4]ICSC − National Research Center in High Performance Computing, Big Data e Quantum Computing, Via Magnanelli 2, 40033 Casalecchio di Reno (BO), Italy; [5]DIATI − Polytechnic University of Turin, Corso Duca degli Abruzzi 24, 10129, Turin, Italy; [6]CNR − Institute of Atmospheric Science and Climate, Corso Fiume 4, 10133, Turin, Italy

* e-mail: erica.bisesi@inaf.it; tel.: +39 040 3199233

## Abstract

Between 2.4 and 0.6 Gy ago, our planet underwent several episodes of global glaciations, including the "Snowball Earth" case that ended 635 My ago. Causes of this last Snowball event presumably included a decreased greenhouse gas concentration and high continental albedo, both associated with the passage of the super-continent Rodinia at equatorial latitudes. When large continental masses are in equatorial regions, silicate weathering is enhanced, leading to decreased atmospheric $CO_2$ concentration, while the bare continental masses, which at the time hosted no vegetation, enhanced reflection of solar radiation. Since then, no other Snowball episodes were recorded. Here we numerically explore the climatic dynamics of a rocky planet for different values of solar output, continental configuration (current and Rodinia-like), $CO_2$ concentration and continental albedo, simulating the effects of land vegetation. We found that for the solar input typical of 600-700 My ago (95% of the current value), the presence of bare continents with albedo 0.35 (granite) in the position estimated for Rodinia was sufficient to trigger a Snowball state for $CO_2$ concentrations up to at least 1000 ppm. When bare continents are located in modern positions, Snowball could be triggered only for values of $CO_2$ concentration below 400 ppm. At current solar input values, Snowball states appear only at or below 100 ppm. Thus, we found that (a) a lower solar output is an essential component of the transition to Snowball; (b) the presence of land vegetation is crucial and reduces the probability of entering a Snowball state; (c) a low $CO_2$ concentration was not needed for triggering a Snowball in bare Rodinia-like conditions and reduced solar output; and (d) current solar luminosity does not allow Snowball states, even for equatorial continents, unless continental albedo is that of granite and $CO_2$ concentration is 100 ppm or less.

**Introduction**

The climate of a rocky planet or satellite having an atmospheric envelope is a complex system regulated by several components and feedback processes (Pierrehumbert 2010). Apart from the external parameters, such as the impinging stellar energy and the orbital characteristics, two of the most important internal control parameters, whose values partially depend on the climate itself and on the characteristics of the biosphere (when present), are the planetary albedo and the intensity of the greenhouse effect. Variations in both the external and internal parameters can lead to large fluctuations in the planet's climate, with the possibility of transitions from one stable equilibrium to another (Dijkstra 2018). In this work, we numerically explore the effects of such parameter variations and the role of the biosphere in the emergence of a globally glaciated state, such as the Snowball Earth episode at about 635 million years ago (Harland 1964, Kirschivink 1992).

The albedo measures the reflectivity of the planet's surface and atmosphere, and it is defined as the fraction of the incoming stellar radiation that is reflected into space, without entering the thermodynamical machinery of the climate (Pierrehumbert 2010). White or light surfaces, such as snow, ice, clouds, bare granite and deserts, have high albedo for impinging light in the visible spectrum, while dark surfaces, such as forests and oceans have (much) lower albedo. The higher the albedo, the less stellar energy is absorbed and the lower is the surface temperature of the planet; currently, on average the Earth is reflecting about 30% of the received solar radiation.

The greenhouse effect, instead, modulates the amount of (usually infrared) radiation emitted by the planetary surface heated by the absorbed stellar radiation. This emitted radiation, however, is not flowing freely to space, but it is partially absorbed by the atmosphere and some of it is re-emitted towards the surface (Pierrehumbert 2010). On Earth, the atmospheric absorption of infrared radiation from the planet's surface depends on the presence of trace gases such as methane ($CH_4$) and carbon dioxide ($CO_2$) and on water vapour. The larger is the concentration of such trace constituents, the stronger is the greenhouse effect and the higher is the surface temperature. Both these effects act in parallel to determine the climate, together with atmospheric and oceanic (when present) transport mechanisms which redistribute heat across the planet. Just to make an example, in current Earth conditions the greenhouse effects is able to raise the planet's surface temperature from about -18 °C (roughly the value which would be attained with no atmosphere) to about 15 °C.

Between about 2.4 and 0.6 billion years ago, our planet has undergone several episodes of large-scale glaciations, including the last and best-known one that ended about 635 million years ago, just before the beginning of the Cambrian Period, known as "Snowball Earth" (Kirschivink 1992, Hoffman & Schrag 2002). In reality, it is not yet clear whether the ice cover was truly global or whether there was a belt of open ocean waters in contact with the atmosphere around the Equator, in the so-called "Slushball" or "Waterbelt" state (Crowley et al. 2001). In any case, one of the processes that allowed the persistence of a glaciated state was the positive feedback between ice and albedo: when temperature drops, for any reason, ice starts to expand. The higher albedo of the ice leads to still lower temperatures, which in turn lead to the formation of more ice, and so on till a global glaciation is established (Budyko 1969; Sellers 1969). Simple climate models show that for many values of the stellar radiation input in the general range of solar luminosity, two stable states can co-exist: a Snowball equilibrium and a "temperate" state with limited ice cover (North et al. 1981). Recent work has shown that such bi-stability is found for values of the stellar, orbital and climatic parameters that correspond, quite closely, to the "habitable" region of parameter space, in the sense of the presence of liquid water at the planet's surface (Murante et al. 2020).

While the ice-albedo feedback is known since long time, the precise causes that can trigger a global glaciation are still debated. First, the solar luminosity was lower than today (Schwarzschild 1958).

For the specific case of the last Snowball episodes, between about 700 and 635 million years ago (so-called Sturtian and Marinoan glaciations, see e.g., Arnaud et al. 2011) the solar luminosity was about 95% of the current value. In addition, owing to the irregular motions generated by plate tectonics, the super-continent Rodinia was located at equatorial and tropical latitudes, and it was fragmenting into smaller blocks with large expanses of continental shelf and internal shallow and warm oceans (Pisarevsky et al. 2003). This geographical setting brought with it an important effect: rock weathering was enhanced by the fact that the continental surfaces were exposed to the warmth and to the strong precipitations of the low latitudes. As it is well-known from the geological carbon cycle, rock weathering and the subsequent burial of carbonates in the continental shelf leads to a decrease of the atmospheric $CO_2$ and, consequently, of the greenhouse effect (Wallmann & Aloisi 2012).

Another very important point is that 700 million years ago the continents were essentially devoid of vegetation and the bare granite surfaces had a much larger albedo than oceans or forested surfaces. Since these almost bare granite continents were at equatorial latitudes, where most of the solar radiation was received, the reflection of sunlight was very significant. Both effects may have contributed to a decrease the surface temperature, favouring the triggering of a global glaciation. When the ice reached low enough latitudes, the high albedo of ices and snow accelerated the process through the ice-albedo feedback and the Earth fell into a Snowball state. In the following hundreds of thousands of years, the volcanoes continued to emit carbon dioxide and methane and after sufficient time the concentration of such gases, together with other effects presumably related to the properties of the cloud cover, generated enough greenhouse effect to melt the Snowball and led to the planet as we now know it (Hoffman et al. 1998).

In the last 600 million years, the Earth did not undergo any further Snowball state, even though the continents were close to the Equator at times. One could then wonder what had changed in the triggering mechanism since pre-Cambrian times. Along these lines, one important thing happened: vegetation established itself on continental surfaces (e.g., Morris et al. 2018), transforming them from bare rocks to forests and grasslands. In doing so, vegetation heavily changed the albedo of the continents, from the higher values of granite (of about 0.35) to the typical values for forests, around 0.15. Under these conditions, a different process can become active: the vegetation-albedo feedback, known as the "Charney mechanism" (Charney et al. 1975; Baudena et al. 2009; Cresto Aleina et al. 2013).

Since vegetation is usually darker than bare rock or sand, its lower albedo tends to locally warm up the surface and increase atmospheric instability, convection and thus precipitation. In this way, vegetation can enhance the hydrological cycle and, in arid or semi-arid land where vegetation growth is limited by water availability, supports its own persistence. Although there are other important effects that complicate the picture, such as the role of vegetation in enhancing evapotranspiration and carbon removal from the atmosphere (e.g., Baudena et al. 2009), which can both reduce the warming effect of vegetation, the vegetation-albedo feedback is a relevant mechanism that can lead to multiple stable equilibria characterized by a different temperature and vegetation cover.

In a previous study (Bisesi et al. 2024), we updated the `ESTM` (`Earth-Like Surface Temperature Model`; Vladilo et al. 2015; Silva et al. 2017, 2023) by incorporating a new set of differential equations to account for the presence, distribution and competitive dynamics of diverse vegetation types—specifically grasses and trees (the latter modeled in both seedling and adult stages). This allowed us to quantify the impact of vegetation albedo on the habitability of Earth-like exoplanets. As a consequence of a decrease in surface albedo, we estimated how vegetation was able to extend the planet's circumstellar habitable zone beyond its external border.

In this work we explore the dynamics of rocky planets that do or do not harbour vegetation on land, and study the dynamics in different continental configurations including the current Earth and the case of equatorial Rodinia about 700 million years ago, for different values of greenhouse gas concentrations in the atmosphere. We explore what parameter values and planetary characteristics are compatible with the insurgence of a Snowball state, and when, instead, such a glaciated state is difficult to trigger. We believe that the results of this study provide further and relevant information on the conditions leading to Snowball states in rocky planets.

**Aims**

Before the Cambrian, continents on Earth did not have a vegetation cover and their albedo could have been as high as that of granite (0.35) (Kirschivink 1992). On the other hand, vegetation albedo can significantly affect Earth's climate via the Charney mechanism (Charney 1975; Bisesi et al. 2024). The question, then, is whether the presence of land vegetation can prevent the onset of Snowball phases by its effect on surface albedo, considering different values of $CO_2$ concentration in the atmosphere. The aim of this work is to explore this issue by means of ideal numerical experiments with a simplified climate model (extension of that described by Vladilo et al. 2015), allowing a vegetation cover on the Earth's continents, and verify under what conditions a Snowball phase can happen.

**Methods**

Climate model. To investigate the potential role of surface albedo in influencing the onset of Snowball Earth conditions, we employed the Earth-like Surface Temperature Model (`ESTM v3.5`, Vladilo et al. 2015; Silva et al. 2023; Bisesi et al. 2024). `ESTM` is a latitudinal-seasonal energy-balance model (`EBM`), a class of models that are particularly suited for exploring the effects of different climate forcing factors and parameters with poorly known values. Based on the diffusive approximation commonly used for this class of models (e.g., Williams & Kasting 1997; Spiegel et al. 2008), `ESTM` simulates the thermal state of a planet by solving a modified diffusion equation:

$$C\frac{\partial T}{\partial t} - \frac{\partial}{\partial x}\left[D(1-x^2)\frac{\partial T}{\partial x}\right] = [S(1-A)] - I \qquad (1)$$

where $t$ denotes time, $x = \sin(\theta)$ is the sine of the latitude and $T = T(t, x)$ the corresponding latitudinal band temperature. This approach accounts for both the latitudinal heat transport, modulated by the diffusion coefficient $D$, and the energy exchange between the surface, atmosphere, and space. The model explicitly calculates the effective thermal capacity, $C$, by weighting the contributions of different surface types (land, ocean, ice over lands, and ice over oceans) according to their zonal coverage. The stellar forcing term, $S = S(t, \theta)$, incorporates orbital parameters and axial tilt to determine the incident radiation. A key feature of the model is its treatment of the Top-of-Atmosphere (TOA) albedo, $A$, and the Outgoing Long-wave Radiation (OLR), $I$. Both $I$ and $A$ are computed through single-column radiative transfer calculation. In addition to the temperature dependence, $A$ is also determined by the distribution of surface components and cloud cover. To ensure physical grounding under varying atmospheric conditions, the climate computation has been coupled with pre-computed radiative transfer (RT) look-up tables. These tables are interpolated as a function of temperature, surface albedo, zenith angle, atmospheric pressure, composition, and planetary gravity. This hybrid approach enables `ESTM` to efficiently explore the vast climatic parameter space through physically grounded simulations of surface temperatures across a diverse range of planetary and surface conditions. Ice cover is treated through a simplified approach: for each latitudinal band, the ice fraction is estimated using a sigmoid function based on the mean temperature of the preceding six months. Our version builds upon the parametrization of Vladilo et al. (2015) and Bisesi et al. (2024), as summarized in Table 1.[1]

Table 1. Physical parameters adopted in our climate simulations.

| Parameter | Value | Comment |
|---|---|---|
| $C_l$ | 0.50 e6 [J m$^{-2}$ K$^{-1}$] | Thermal capacity per unit area of land [a] |
| $C_o$ | 2.10 e8 [J m$^{-2}$ K$^{-1}$] | Thermal capacity per unit area of ocean [a] |
| $a_{sl}$ | [0.10–0.35] | Surface albedo of lands |
| $a_{sil}$ | 0.70 | Surface albedo of ice on land |
| $a_{sio}$ | 0.70 | Surface albedo of ice on ocean |
| $f_{cw}$ | 0.70 | Cloud coverage on water |
| $f_{cl}$ | 0.60 | Cloud coverage on land |
| $f_{ci}$ | 0.60 | Cloud coverage on ice |

**Note.** [a] According to Pierrehumbert 2010 and Williams & Kasting 1997.

The influence of vegetation was investigated by adjusting the fiducial albedo of land surfaces, transitioning from mineral-based values (0.35 for granite; Dobos 2020) to biological ones, ranging from 0.35 down to 0.1 (Bisesi et al. 2024). Note that, as detailed in Vladilo et al. (2015), the ocean albedo is computed as a function of the mean solar zenith angle. For the modern Earth, this yields an average ocean albedo of approximately 0.06. This value is significantly lower than that of granite, and even falls below the albedo of the darkest vegetation considered in this study.

Energy Balance Models (EBMs) in general, and the ESTM in particular, are not intended to provide a physically exhaustive description of the climate system; rather, they serve as idealized frameworks exclusively designed for conceptualizing meridional heat transport. Consequently, the primary relevant output yielded by the ESTM is its asymptotic final state, which, under the influence of periodic forcings such as those dictated by Earth's orbital parameters and axial tilt, manifests inherently as a limit cycle. The transient phase—namely, the progression from the prescribed initial conditions to the asymptotic solution of differential equation Eq. (1)—must not be confused with a realistic climatological evolution. It constitutes merely the mathematical trajectory fundamentally required to converge upon the equilibrium limit cycle. Naturally, this paradigm would shift were the system initialized from a pre-existing equilibrium and subsequently perturbed by a transient forcing during the integration; however, such a scenario falls outside the scope of the simulations presented herein. All simulations were initiated via a 'hot start' protocol, with the global temperature uniformly set to 275 K. During the initial ten orbits, the ice fraction was calculated based on instantaneous temperature values to facilitate rapid convergence. Once the system approached its equilibrium state, the model transitioned to using time-averaged temperatures to ensure long-term stability.

The look-up tables were generated using the software petitRADTRANS (pRT, Mollière et al. 2019), an extremely flexible radiative transfer code, specifically designed to simulate a wide range of atmospheric structures and compositions (Alei et al. 2022, Villanueva et al. 2024). It includes a built-in database of default opacity tables, based on the most widely used repositories—HITRAN, HITEMP, and ExoMOL—but it can be easily interfaced with user-defined opacity sources. In this work, we adopted the default opacity tables provided with pRT, and we refer to the new coupled atmospheric and climate model as `pRT-ESTM`. We have fed the `RT` code with pressure-temperature (*P-T*) profiles derived from moist adiabats, which approximate the atmospheric structure under radiative–convective equilibrium (Pierrehumbert 2010). Consistent with our previous studies, we assumed an atmospheric relative humidity (RH) of 0.6 throughout the troposphere and a constant stratospheric temperature of 200 K. All tables have been obtained using an oxygen partial pressure of 0.15 bar and varying the $CO_2$ concentration from 100 to 10,000 ppm. The remaining atmospheric gas is nitrogen, $N_2$, bringing the total ground-level dry pressure to 1 bar.

---

[1] For an in-depth analysis of the model's framework, including how the variables *C, D, S, A,* and *I* incorporate the planet's astrophysical, geophysical, and atmospheric parameters, the reader is referred to Vladilo et al. (2015).

Surface geography. In the simulations, two different geographic setups were considered: modern Earth and a Rodinia-like continental distributions, in both cases with an ocean fraction of 70% (Fig. 1). The Rodinia ocean fraction as a function of latitude was extrapolated from the reconstruction published by Li et al. (2008); note the equatorial distribution of continents in this case. At low latitudes, especially in the Northern Hemisphere, a significant fraction of the planet's surface is covered by continents, similar to modern Earth geography.

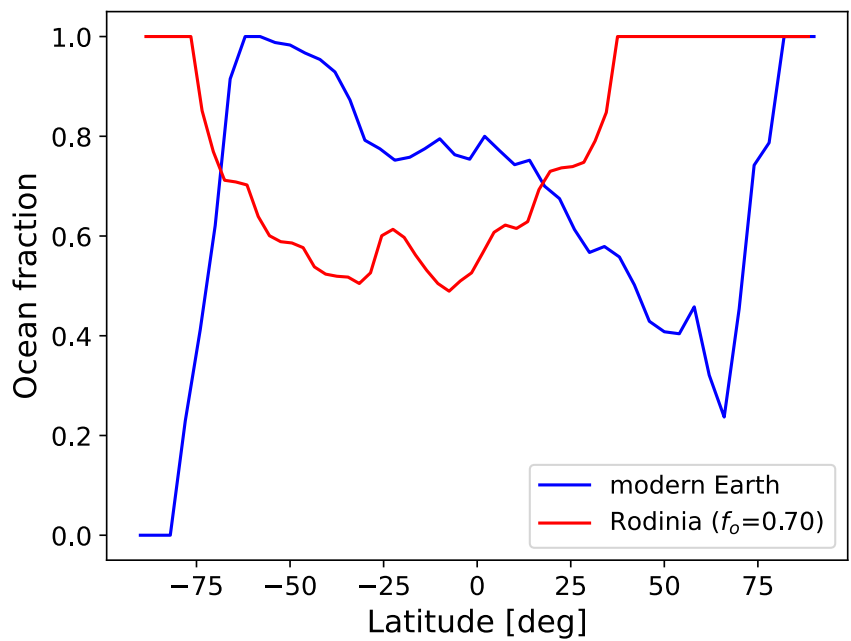

**Figure 1.** Geography (in terms of ocean cover) for the two configurations (modern Earth and Rodinia-like) considered in this work, as a function of the latitude. The parameter $f_o$ indicates the global ocean fraction and it is 70% in both cases.

Insolation configuration. Finally, we considered two different values of the solar luminosity, since in the Precambrian era at the time of the last Snowball, the Sun was fainter than now by about 5% (Kienert et al. 2012). In the simulations, we thus considered the two cases with solar luminosity 0.95 $L_\odot$ and $L_\odot$, where $L_\odot$ is the current value of the solar luminosity, leading to the so-called solar constant for Earth having an average value of 1361 W m$^{-2}$.

**Results**

The main results of the simulations are reported in Fig. 2. In the upper panel, solar luminosity is set to 0.95 $L_\odot$, corresponding to what is currently assumed to be in the Cambrian Earth. On the x-axis, we considered seven different atmospheric compositions, respectively: 100 ppm, 200 ppm, 360 ppm, 400 ppm (roughly corresponding to modern Earth), 1000 ppm, 2000 ppm, and 10000 ppm. On the y-axis, we varied the albedo from that of granite (0.35, upper line), to lower values accounting for different values of vegetation albedo (lighter: 0.20, grass-like; intermediate: 0.15; dark: 0.10, woods-like). Circles refer to modern Earth; diamonds to Rodinia. The color scale indicates the ice fraction:

white (light tones in greyscale) means that the planet is in a Snowball state; yellow/light grey means that the planet is in a Waterbelt state; from orange to red (intermediate gray) correspond to a planet with lower and lower ice fraction; brown/darkest gray corresponds to a planet with no ice at all. As a comparison, in the lower panel we show the results obtained for a modern solar luminosity.

In the Precambrian Earth, at solar luminosity 0.95 $L_\odot$ and $CO_2$ fraction from 100 to 400 ppm, a planet with bare granite continents freezes—independent on whether the geography is that of the modern Earth of Rodinia. At 1000 ppm, only the Rodinia-like continental distribution would end in a Snowball state, while the modern distribution of continents would retain a large surface free of ice. For larger values of the $CO_2$ concentration, the ice cover decreases for both types of continental configuration, leading to a substantially ice-free planet for a $CO_2$ concentration of 10,000 ppm.

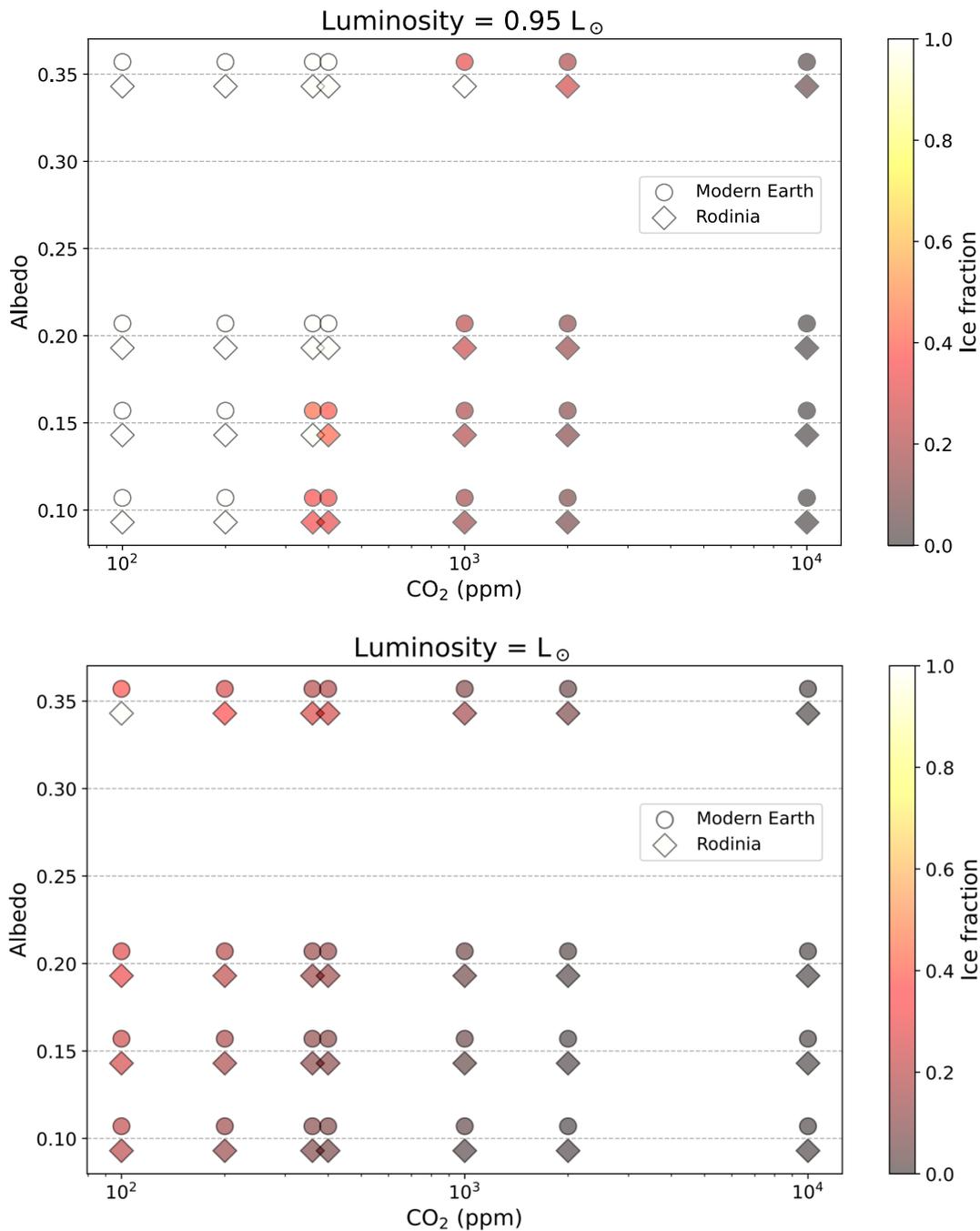

**Figure 2.** Upper panel: Climate type (from Snowball, in white, to fully deglaciated, in brown) for a Cambrian solar luminosity (0.95 $L_\odot$). On the x-axis, the assumed atmospheric carbon dioxide

concentration in ppm; on the y-axis, the assumed value of albedo. Circles: continental distribution as in modern Earth; diamonds: continental distribution as in Rodinia just before the last Snowball Earth episode. Color scale: ice fraction: white (light tones in greyscale), Snowball state; yellow (light grey), strong Waterbelt state; from orange to red (intermediate gray), lower and lower ice fraction; brown (darkest grey): no ice. Lower panel: Climate type for a modern Earth solar luminosity ($L_\odot$), same details as for the upper panel.

By introducing increasingly darker vegetation, the $CO_2$ concentration below which the planet enters a Snowball state shifts to lower values. For a continental albedo value of 0.15, a planet with a Rodinia-like continental distribution shifts from partially ice-free conditions to a Snowball state when the $CO_2$ fraction falls below a threshold between 360 ppm and 400 ppm. For $CO_2$ concentration values of 1000 ppm, both geographic configurations remain in partially ice-free states. For modern values of the solar luminosity, $L_\odot$, only a bare-land granite planet with a $CO_2$ fraction of 100 ppm enters a Snowball state, independent on the continental configuration. By increasing the $CO_2$ fraction and/or by lowering the albedo, it progressively moves to an increasingly hotter and ice-free surface. Note, also, that this simple model indicates that modern Earth would become essentially ice-free when $CO_2$ concentration values are above 1000 ppm.

As discussed above, all primary simulations were initialized with a 'hot start' configuration, meaning they were initially ice-free with a uniform temperature of $T = 275$ K. To investigate potential bistability, we repeated the simulations using a 'cold start' approach, characterized by an initial uniform temperature of 250 K and a globally frozen surface for the first 10 orbits. Under 'cold start' conditions, the planet failed to exit the Snowball state at $L = L_\odot$ in both the Rodinia and modern Earth configurations. In these cases, at least 40,000 ppm of $CO_2$ are required to reach an average surface temperature exceeding 273.15 K, at which point the resulting average temperature consistently exceeds 50 °C. Consequently, at this solar flux value, all cases exhibit bistability, yielding either a warm climate or a Snowball state depending on the initial conditions. This pattern persists for $L = 0.95\,L_\odot$, where 100,000 ppm of $CO_2$ are necessary to deglaciate the planet. Thus, while configurations with lower $CO_2$ concentrations admit only the Snowball solution, all unfrozen states at $L = 0.95\,L_\odot$ are bistable. We note that the $CO_2$ partial pressures required to exit the Snowball state, though high, align with recent literature (e.g., Thomas et al., 2025).

We also examined the influence of obliquity and eccentricity to assess the impact of seasonal variations on surface temperature. To isolate these effects, we performed sensitivity experiments by imposing zero eccentricity and zero axial tilt. Under these conditions, both the Rodinia and modern Earth configurations yield warmer climates compared to the baseline seasonal scenarios. Notably, the single Snowball state observed at $L = L_\odot$ disappears entirely. At $L = 0.95\,L_\odot$, Snowball states persist only under specific conditions: at 100 ppm of $CO_2$ (with an albedo above 0.1), at 200 ppm with bare continents, and at 360 ppm specifically for the Rodinia bare-continent geography. These results suggest that seasonal ice advance toward lower latitudes plays a critical role in triggering the runaway ice-albedo feedback loop, and thus that the values of eccentricity and obliquity are important to determine whether the climate system will end up in a Snowball state.

**Discussion**

The triggering of Snowball states in rocky planets requires an initial temperature decrease followed by a sequence of feedback effects, eventually leading to the transition to an equilibrium different from the "temperate" one. In the history of Earth, there are evidences of several occasions in which the planet was covered with large expanses of ice, either in terms of a truly global ice cover or a so-called "Waterbelt" state in which the low-latitude portions of the ocean were at least partially free of ice (Kirschivink 1992, Crowley et al. 2001). The first known occurrence is the Huronian glaciation, dated about 2.4 billion years ago, that took place soon after the rise of atmospheric oxygen (Coleman 1907). Another well-known event developed between about 700 and 635 million years ago, specifically known as the "Snowball Earth" (Kirschivink 1992, Hoffman et al. 1998). Previous exploration using the simplified climate model adopted here showed that the bistability between a temperate and a Snowball state is rather common in the planetary and climatic parameter range where liquid water is present at the surface and the planet is habitable (Murante et al. 2020), thus suggesting that Snowball occurrences are not a rare event in the life of potentially habitable rocky planets.

The causes of these extended, possibly global glaciations are varied and still debated, but in general they involve two main mechanisms: a decrease of the absorbed stellar radiation, due to an increase of the albedo, and a decrease of the greenhouse effect due to changes in atmospheric composition (Hoffman & Schrag 2002). In the case of the Huronian glaciation, it is thought that the main trigger was the decrease of atmospheric methane owing to the increase of oxygen, which reacted with methane producing carbon dioxide and water (Canfield 2015). Since both carbon dioxide and water vapour are less efficient than methane as greenhouse gases, a strong temperature decrease could have followed and the planet plunged into a Snowball state. In the case of the last event, on the other hand, it is thought that the main causes were related to the presence of the large continental masses of the super-continent Rodinia at tropical and equatorial latitudes, with the double effect that (a) rock weathering was enhanced and consequently carbon dioxide removal from the atmosphere was stronger, and (b) the largely bare continental surfaces reflected a larger proportion of the incoming solar radiation. In addition, in all those cases the incoming solar radiation was smaller, reaching at most 95% of the current value during the last Snowball episode.

Here, we focused on the role of the position of continental masses, with its effect on albedo and greenhouse effect, ideally focusing on the last Snowball event. Using the `pRT-ESTM` simplified climate model, we showed that for the solar input typical of 600-700 million years ago (95% of the current value), the presence of bare continents with mean albedo 0.35 in the position estimated for Rodinia was sufficient to trigger a global Snowball state for $CO_2$ concentrations up to at least 1000 ppm. If continents were instead located in the modern positions, Snowball could be triggered only for values of $CO_2$ concentration up to 400 ppm. At current values of solar input, instead, Snowball states can appear only at or below 100 ppm with an Equatorial continent distribution like that of Rodinia. For $CO_2$ partial pressures larger than 200 ppm, no extended icy areas are present. For $CO_2$ concentration around 200 ppm and bare continents, extended icy areas can still be present, with ice-covered planet fractions between about 0.7 for Rodinia-like land distribution and 0.3 for the modern position of the continents.

The presence of vegetation over the continents, on the other hand, modifies the average land albedo and completely changes the situation. For a solar luminosity of 95% of the current value, an average albedo of 0.15 keeps the planet out of Snowball for $CO_2$ concentration values of 400 ppm and higher, with this threshold moving to greater values for increasing albedo and lower values for lower albedo (e.g., it is between 200 and 360 ppm for land albedo 0.10). For albedo values between 0.15 and 0.20 there is a rapid increase of the minimum value of $CO_2$ concentration needed to keep the planet out of Snowball. For the current position of the continents, Snowball states are slightly less favoured, as expected. For solar luminosity at current values, the only occurrence of Snowball states (Rodinia-like continental positions) is observed for land albedo 0.35 (bare continents) and $CO_2$ concentration of 100 ppm or lower, indicating that a transition to Snowball is extremely difficult under these conditions.

Since the last occurrence about 635 million years ago, no other Snowball states were recorded on Earth, although glacial-temperate oscillations continued to happen. In the period after the last

Snowball, life bloomed on Earth and started to populate the continents, changing the land albedo. The results of our analysis indicate that the value of land albedo, associated with the presence of vegetation, played a major role in avoiding the trigger of Snowball, which would have required much lower values of $CO_2$ concentration than presumably present at those times. The increase of solar luminosity amplified this effect, making a Snowball even less probable. The different distribution of the continents helped further, limiting the $CO_2$ concentration decrease due to the amplified surface weathering at low latitudes. Our analysis indicates also that seasonal ice expansion toward low latitudes is a key trigger of runaway ice-albedo feedback, so eccentricity and obliquity critically control the likelihood of Snowball onset.

To complete the picture, future analyses will have to consider the full picture of how vegetation interacts with climate. Plants affect the water cycle through evapotranspiration, amplifying water vapour release to the atmosphere from continental masses with respect to pure evaporation from bare soil (e.g., Porporato 2022). Vegetation and soil microbiota participate in the biological carbon cycle, modulating the carbon balance through photosynthesis, respiration and decomposition of organic matter (Williamson et al. 2006). Notably, the relative importance of all these mechanisms depends on both the prevailing climatic conditions and the relevant temporal scales. Given that the timescales of these processes mostly align with those of vegetation dynamics—ranging from decades to millennia—the feedbacks associated with the fast biological carbon cycle can be effectively investigated using an extension of the modeling framework adopted here.

In a longer-term perspective, the `pRT-ESTM` model can also be adapted to incorporate the much slower carbonate-silicate cycle, or "geological carbon cycle", which operates on timescales of tens to hundreds of millions of years. Both processes—the fast and slow carbon cycles—alter the atmospheric $CO_2$ levels, thereby regulating the global greenhouse effect. Within this framework, the objective would then be to characterize the synergistic interactions between these cycles and how they are influenced by planetary and astrophysical parameters—specifically in the contexts of early Earth, early Mars, and terrestrial exoplanets.

**Conclusions**

We used the `pRT-ESTM` climate model, obtained as an extension of the `ESTM` model of Vladilo et al. (2015) coupled with atmospheric column radiative-convective calculations performed with `petitRADTRANS` (Mollière et al. 2019), to quantitatively explore what conditions can trigger the emergence of a Snowball state, ideally focusing on the last known Snowball episodes between about 700 and 635 million years ago. We considered in particular the role of vegetation-induced land albedo changes together with the effect of varying atmospheric $CO_2$ concentration, continental mass distribution and solar luminosity values.

We found that a vegetation cover decreasing the land albedo from 0.35 to 0.15 or lower is sufficient to prevent the onset of a Snowball Earth state for $CO_2$ concentration values of at least 400 ppm, assuming a solar luminosity of 95% the current value and a Rodinia-like distribution of the continents. Modern solar luminosity values do not allow Snowball states, even with equatorial continents, unless the continental albedo is close to that of granite and $CO_2$ concentration is 100 ppm or less.

Triggering a Snowball state thus depends on a multitude of factors, including the amount of incoming stellar radiation, the atmospheric greenhouse gas concentration, the position of the continents and the land albedo, and seasonality. The results presented here underscore the pivotal role of vegetation in shaping a planet's climatic state, reinforcing our previous findings on the interplay between biosphere distribution and planetary habitability (Bisesi et al. 2024). As a final note, we acknowledge that the present study did not explore the role of vegetation in modulating the carbon cycle—specifically regarding enhanced $CO_2$ uptake—nor its influence on the hydrological cycle through increased transpiration relative to bare-soil evaporation. Future developments of this modeling framework will aim to incorporate these biogeochemical and biophysical feedbacks to provide a more comprehensive representation of planetary habitability.

**Acknowledgements**

This work was supported by the CNR-IGG and by the Italian Space Agency with the `ASTERIA` (ASI No. 2023-5-U.0) project and, partly, by the INAF RSN2 `ClimHAB-RBA` (`Climate Habitability: Refraction and Biosignatures in Habitable Exoplanet Atmospheres`) MINI-GRANT No. 1.05.12.04.02. We sincerely thank the anonymous referee for the insightful comments and constructive feedback provided.

**References**

Alei, E., Konrad, B.S., Angerhausen, D., Grenfell, J.L., Mollière, P., Quanz, S.P., Rugheimer, S., Wunderlich, F. 2022, Large Interferometer for Exoplanets (LIFE), *A&A*, **665**, A106; https://doi.org/10.1051/0004-6361/202243760

Arnaud, E., Halverson, G.P., Shields-Zhou, G.A. 2011, The Geological Record of Neoproterozoic Glaciations, *Geological Society of London, Memoirs*, **36**, 1-16; https://doi.org/10.1144/M36

Baudena, M., D'Andrea, F., Provenzale, A. 2009, An Idealized Model for Tree-Grass Coexistence in Savannas: The Role of Life Stage Structure and Fire Disturbances, *J. Ecol., 98(1)*, 74-80; https://doi.org/10.1111/j.1365-2745.2009.01588.x

Bisesi, E., Murante, G., Provenzale, A., Biasiotti, L., Hardenberg, J.v., Ivanovski, S., Maris, M., Monai, S., Silva, L., Simonetti, P., Vladilo, G. 2024, Impact of Vegetation Albedo on the Habitability of Earth-Like Exoplanets, *MNRAS*, **534**, 1-11; https://doi.org/10.1093/mnras/stae2016

Budyko, M.I. 1969, The effect of Solar Radiation Variations on the Climate of the Earth, *Tellus*, **5**, 611-619; https://doi.org/10.1111/j.2153-3490.1969.tb00466.x

Canfield, D.E. 2015, *Oxygen: A Four Billion Year History*, Princeton University Press; ISBN: 978-0-691-14502-0

Charney, J., Stone, P.H., Quirk, W.J. 1975, Drought in the Sahara: A Biogeophysical Feedback Mechanism, *Science*, **187**, 434-435; https://doi.org/10.1126/science.187.4175.434

Coleman, A.P. 1907, A Lower Huronian Ice Age, *American J. of Science*, **23**, 187-192; https://doi.org/10.2475/ajs.s4-23.135.187

Cresto Aleina, F., Baudena, M., D'Andrea, F., Provenzale, A. 2013, Multiple Equilibria on Planet Dune: Climate-Vegetation Dynamics on a Sandy Planet, *Tellus, 65(1)*, 17662-17673; https://doi.org/10.3402/tellusb.v65i0.17662

Crowley, T.J., Hyde, W.T., Peltier, W.R 2001, $CO_2$ Levels Required for Deglaciation of a "Near-Snowball" Earth, *Geophys. Res. Lett*. **28**, 283-286; https://doi.org/10.1029/2000GL011836

Dijkstra, H.A. 2018, Multiple Equilibria in the Climate System, in Oxford Research Encyclopedia of Climate Science, https://doi.org/10.1093/acrefore/9780190228620.013.85

Dobos E., 2020, Albedo, in Y. Wang, *Atmosphere and Climate*, CRC Press, Chicago, pp. 25-27; ISBN 978-0-429-44565-1

Harland, W.B. 1964, Critical Evidence for a Great Infra-Cambrian Glaciation, *Geologische Rundschau*, **54**, 45-61; https://doi.org/10.1007/BF01821169

Hoffman, P.F., Kaufman, A.J., Halverson, G.P., Schrag, D.P. 1998, A Neoproterozoic Snowball Earth, *Science*, **281**, 1342-1346; https://doi.org/10.1126/science.281.5381.1342

Hoffman, P.F. and Schrag, D.P. 2002, The Snowball Earth Hypothesis: Testing the Limits of Global Change, *Terra Nova*, **14,** 129-155; https://doi.org/10.1046/j.1365-3121.2002.00408.x

Kienert, H. Feulner, G., Petoukhov, V. 2012, Faint Young Sun Problem More Severe Due to Ice-Albedo Feedback and Higher Rotation Rate of the Early Earth, *Geophys. Res. Lett.*, **39**, L23710-L23713; https://doi.org/10.1029/2012GL054381

Kirschivink, J.L. 1992, Late Proterozoic Low-Latitude Global Glaciation: The Snowball Earth, in J.W. Schopf and C. Klein, *The Proterozoic Biosphere. A Multidisciplinary Study*, Cambridge University Press, pp. 51-52; ISBN: 9780521366151

Li, Z.X., Bogdanova, S.V., Collins, A.S., Davidson, A., De Waele, B., Ernst, R.E., Fitzsimons, I.C.W., Fuck, R.A, Gladkochub, D.P., Jacobs, J., Karlstrom, K.E., Lu, S., Natapov, L.M., Pease, V., Pisarevsky, S., Thrane, K., Vernikovsky, V. 2008, Assembly, Configuration, and Break-Up History of Rodinia: A Synthesis, *Precambrian Res*., **160***(1/2)*, 179-210; https://doi.org/10.1016/j.precamres.2007.04.021

Mollière, P., Wardenier, J.P., van Boekel, R., Henning, Th., Molaverdikhani, K., Snellen, I.A.G 2019, petitRADTRANS. A Python Radiative Transfer Package for Exoplanet Characterization and Retrieval, *A&A,* **627**, A67; https://doi.org/10.1051/0004-6361/201935470

Morris, J.L., Puttick, M.N., Clark, J.W., Edwards, D., Kenrick, P., Pressel, S., Wellman, C.H., Yang, Z., Schneider, H., Donoghue, P.C.J. 2018, The Timescale of Early Land Plant Evolution, *PNAS*, **115**, E2274-E2283; https://doi.org/10.1073/pnas.1719588115

Murante, G., Provenzale, A., Vladilo, G., Taffoni, G., Silva, L., Palazzi, E., Hardenberg, J.v., Maris, M., Londero, E., Knapic, C., Zorba, S. 2020, Climate Bistability of Earth-Like Exoplanet, *MNRAS*, **492**, 2638-2650; https://doi.org/10.1093/mnras/stz3529

North, G., Cahalan, R., Coakley, J. 1981, Energy Balance Climate Models, *Rev. Geophysics and Space Physics*, **19**, 91-121; https://doi.org/10.1029/RG019i001p00091

Pierrehumbert, R. 2010, *Principles of Planetary Climates*, Cambridge University Press; ISBN: 978-0-521-86556-2

Pisarevsky, S.A., Wingate, M.T.D., Powell, C., Johnson, S., Evans, D.A.D. 2003, Models of Rodinia Assembly and Fragmentation, in M. Yoshida, B.F. Windley and S. Dasgupta, S., *Proterozoic East Gondwana: Supercontinent Assembly and Breakup*, Geological Society, London, Special Publications, **206**, 35-55; ISBN: 978-1-862-39125-3

Porporato, A. 2022, *Ecohydrology. Dynamics of Life and Water in the Critical Zone*, Cambridge University Press; ISBN: 978-1-108-84054-5

Schwarzschild, M. 1958, *Structure and Evolution of the Stars*, Princeton University Press. ISBN: 978-0-691-65283-2

Sellers, W. 1969, A Global Climatic Model Based on the Energy Balance of the Earth-Atmosphere System, *J. Appl. Meteor.*, **8**, 392-400; https://doi.org/10.1175/1520-0450(1969)008<0392:AGCMBO>2.0.CO;2

Silva, L., Vladilo, G., Murante, G., Provenzale, A. 2017, Quantitative Estimates of the Surface Habitability of Kepler-452b, *MNRAS,* **470***,* 2270-2282; https://doi.org/10.1093/mnras/stx1396

Silva, L., Bevilacqua, R., Biasiotti, L., Bisesi, E., Ivanovski, S.L., Maris, M., Monai, S., Murante, G., Simonetti, P., Vladilo, G. 2023, Climate and Atmospheric Models of Rocky Planets: Habitability and Observational Properties, *Mem. S.A.It.,* **94,** 203-210; https://doi.org/10.36116/MEMSAIT_94N2.2023.203

Spiegel, D.S., Menou, K., Scharf, C.A. 2008, Habitable Climates, *ApJ*, **681**, 1609-1623; https://doi.org/10.1086/588089

Thomas, T. B., Macdonald, F. A., Catling; D. C. 2025, Seafloor Weathering Can Explain the Disparate Durations of Snowball Glaciations, *Geology*; **54**(2), 158-162; https://doi.org/10.1130/G53722.1

Villanueva, G.L., Fauchez, T.J., Kofman, V., Alei, E., Lee, E.K.H., Janin, E., Himes, M.D., Leconte, J., Leung, M., Faggi, S., Mak, M.T., Sergeev, D.E., Kozakis, T., Manners, J., Mayne, N., Schwieterman, E.W., Howe, A.R., Batalha, N. 2024, Modeling Atmospheric Lines by the Exoplanet Community (MALBEC) Version 1.0: A CUISINES Radiative Transfer Intercomparison Project, *Planet. Sci. J.*, **5***(3)*, 64-83; https://doi.org/10.3929/ethz-b-000666020

Vladilo, G., Silva, L., Murante, G., Filippi, L., Provenzale, A. 2015, Modeling the Surface Temperature of Earth-Like Planets, *ApJ,* **804**, 50-69; https://doi.org/10.1088/0004-637X/804/1/50

Wallmann, K. and Aloisi, G. 2012, The Global Carbon Cycle: Geological Processes, in A.H. Knoll, D.E. Canfield and K.O. Konhauser, *Fundamentals of Geobiology*, pp. 20-35; https://doi.org/10.1002/9781118280874.ch3

Williams, D.M. and Kasting J.F. 1997, Habitable Planets with High Obliquities, *Icarus*, **129**, 254-267; https://doi.org/10.1006/icar.1997.5759

Williamson, M.S., Lenton, T.M., Shepherd, J.G., Edwards, N.R 2006, An Efficient Numerical Terrestrial Scheme (ENTS) for Earth System Modelling*, Ecol. Model.,* **198,** 362-374; https://doi.org/10.1016/j.ecolmodel.2006.05.027